\documentclass[10pt]{iopart}

\usepackage{epsf}

\begin{document}

\title{Inherent Structures, Configurational Entropy and 
       Slow Glassy Dynamics}

\author{A. Crisanti\dag\ and F. Ritort\ddag}

\address{\dag\ Dipartimento di Fisica, Universit\`a di Roma ``La Sapienza''
         and \\
         Istituto Nazionale Fisica della Materia, Unit\`a di Roma\\
         P.le Aldo Moro 2, I-00185 Roma, Italy}

\address{\dag\ Physics Department, Faculty of Physics \\
         University of Barcelona, Diagonal 647, 08028 Barcelona,  Spain
        }

\begin{abstract}
We give a short introduction to the inherent structure approach, with
particular emphasis on the Stillinger and Weber decomposition, of
glassy systems. We present some of the results obtained in the
framework of spin-glass models and Lennard-Jones glasses. We discuss
how to generalize the standard Stillinger and Weber approach by
including the entropy of inherent structures. Finally we discuss why
this approach is probably insufficient to describe the behavior of
some kinetically constrained models.

\end{abstract}

\pacs{64.70.Pf, 
      75.10.Nr, 
      61.20.Gy, 
      82.20.Wt}  

\maketitle

\section*{Introduction}
If we are asked what is a glass most probably we will think about a
window-glass. The glassy state is, however, rather common in nature
and many, apparently unrelated, systems such as structural glasses,
spin glasses, disordered and granular materials or proteins among
others, presents what is called a {\it glassy behaviour}.  All these
systems have as common feature a dramatic slowing down of the
equilibration processes when some control parameter, e.g., the
temperature or density, is varied. The equilibration process is
frequently non-exponential, and correlation functions show power-law
and stretched exponential behavior as opposed to a simple exponential
decay.  As the characteristic relaxation time may change of
several orders of magnitude, it easily exceeds the observation
time. The residual very slow motion leads to a non-equilibrium
phenomena which changes the properties of the systems, a process
commonly called aging.  The greatest difficulty in understanding the
slow relaxation glassy dynamics is that a general non-equilibrium
theory to deal with this class of systems is still missing and
approximations to this problem remain partial. They usually work
either in a limited range of time scales or in a limited range of
temperatures (for instance, mode-coupling theory \cite{MCT}).

Following the ideas formulated more than 30 years ago by Goldstein \cite{GO}, 
a convenient framework for understanding the complex phenomenology
of glassy systems is provided by the energy landscape analysis. 
The trajectory of the
representative point in the configuration space can be viewed as a
path in a multidimensional potential energy surface. The
dynamics is therefore strongly influenced by the topography of the
potential energy landscape: local minima, barriers, basins of
attraction an other topological properties all influence the dynamics.

The idea of Goldstein, formulated at a qualitative level, was 
formalized years later by Stillinger and Weber (SW) \cite{SW82,S95}, 
who proposed an operative method to identify basins of the potential 
energy surface of super-cooled liquids.
The recipe is rather simple: 
the set of all configurations connected to the same 
local energy minimum by a steepest-descent path uniquely defines
the basin associated with this minimum, which Stillinger and Weber
named {\it Inherent Structure} (IS) to stress its intrinsic nature. 

The physical motivation behind their proposal follows from the
observation that the potential energy surface of a 
super-cooled liquid contains a large number of local minima.
Therefore the
time evolution of the system can be 
seen as the result of two
different processes: thermal relaxation into basins
({\it intra-basin} motion)  and 
thermally activated potential energy barrier crossing between different
basins ({\it inter-basin} motion). 
When the temperature is lowered down to the order of the critical
Mode Coupling Theory (MCT) temperature $T_{MCT}$ 
the typical barrier height is of the order of the
thermal energy $k_{B} T_{MCT}$, and 
the inter-basin motion slows dominating the 
relaxation dynamics.
If the temperature is further reduced the
relaxation time eventually becomes of the same order of the 
physical observation time and the system falls out of equilibrium since 
there is not enough time to cross barriers and equilibrate. 
This define the {\it experimental glass transition} temperature $T_g$.

With this picture in mind it is natural to view the IS as the
natural elements to describe the slow glassy dynamics. If we think 
of the glassy systems as a dynamical system, then the SW decomposition
is a mapping of the true dynamics at a given temperature
onto the IS dynamics. 
This approach is rather appealing since leads naturally to 
universality: all glassy systems with similar IS dynamics must have 
similar glassy behaviour. 

The increase of computational power has significantly improved the
analysis of the energy surface and IS analysis of the
energy surface has been done for several systems.
The results are both positive and negative, indeed while
the IS formalism has been successful for the 
description of the off-equilibrium dynamics and 
the FDT violations in structural glass 
models\cite{ISok,CR00a,CR00b,CR00,ST01}, 
it fails for some kinetically constrained
glassy systems\cite{ISno}.

To understand this success/failure 
we have to analyse the idea behind the SW approach.
The main question we would like to answer is: {\it what is
a good description for the long time slow glassy dynamics?}
The natural approach is to look for some ``reduced dynamics''
which includes only those details of the full dynamics
relevant on the long time scales.
This obviously implies a coarse graining of the phase space.
For example within the Mori's projectors method used to derive
the MCT, the phase space is coarse grained by projecting it onto the
the subspace spanned by the ``slow variables''.
It is clear that even if the phase space can be always coarse grained, not
all possible coarse graining will lead to a relevant reduced dynamics.
This is a well known problem in the theory of dynamical systems, where
the associated reduced dynamics is called {\it symbolic dynamics}.
Indeed from the theory of dynamical systems we know that a 
symbolic dynamics gives a good description of the full 
dynamics only if the mapping between the full phase space and the 
coarse grained one defines what is called a 
{\it generating partition}, see e.g. Ref. \cite{BS}. 
In general for a generic systems not only it is not trivial to demonstrate that
a generating partition exists but, even when it does exist, 
its practical identification remains a highly non-trivial task so that
we can answer to this question only at posteriori.
We first define a partition and then check if this reproduces
the desired features of the dynamics. 

The SW mapping identifies configuration in a IS-basin with the IS
itself. Therefore to be a plausible mapping the systems must spend a 
lot of time inside the basin. Under this assumption 
the dynamics on time scales larger than the typical residence time inside 
a IS-basin could be quite well described by the IS dynamics. 
This scenario is typical, e.g., of a many valley energy landscape with 
activated dynamics. This, however, is only one of the plausibility 
conditions since other requirements on the dynamics must be satisfied, as 
discussed later in this work.

To illustrate our discussion we shall report results from numerical simulations
of {\it finite-size} fully-connected Ising-spin Random
Orthogonal Model (ROM) \cite{ROM} and binary mixture of 
Lennard-Jones (BMLJ) \cite{KOB}. 
The first one,  belonging to the $p$-spin class \cite{psp}, 
is a fully connect Ising spin glass model with a random 
orthogonal interaction matrix
whose high-temperature dynamics is
described in the thermodynamic limit by the MCT
\cite{REVIEW-DYN}. The second one,
a typical system used for studying the structural glass transition,
is a system composed by a mixture of type $A$ and type $B$ particles 
interacting via a Lennard-Jones potential. One of the main advantages is that
with a suitable choice of the Lennard-Jones potential parameters for
$AA$, $AB$ and $BB$ interactions the 
system does not crystallize simplifying the analysis of the glass
transition.

\section{The Stillinger and Weber decomposition}
\subsection{The SW configurational Entropy}
The recipe of the Stillinger and Weber decomposition 
is rather simple \cite{SW82}: 
the set of all configurations connected to the same 
local energy minimum (IS) by a steepest-descent path uniquely defines
the basin associated with the minimum.
The phase space is then partitioned into a disjoint set of basins, 
usually labeled by $e_{\rm IS}$ the energy of the IS.
Under broad assumptions, e.g., that boundaries between basins are
sub-extensive, this decomposition covers almost all the energy surface
and, collecting  all IS with the same energy, the partition sum is written as
sum of basin partition functions:
\begin{equation}
\label{eq:part}
  Z_N(T)\simeq \int d e \exp\, N\,\left[ s_c(e)
                                     -\beta f_b(T,e)
                               \right]
\end{equation}
where $N\,s_c(e)$ account for the entropic contribution arising
from the number of basins with energy $e_{\rm IS} = e$.
We shall call $s_c(e)$ the
{\it SW configurational entropy or complexity} to distinguish it from other 
possible definitions taken from mean-field  concepts \cite{MG,CMPV}. 
The term $f_b(T,e)$ is the typical free energy 
of the $e_{\rm IS}$ basins with energy $e_{\rm IS} = e$.
If all $e_{\rm IS} = e$ basins have similar statistical properties
then $f_b(T,e)$ is the free energy of the system when constrained in 
one characteristic $e_{\rm IS}$ basin with $e_{IS} = e$. 

In equilibrium at each temperature $T = 1/\beta$ the system will 
visit $e_{\rm IS}$ basin with probability, see eq. (\ref{eq:part}),
\begin{equation}
\label{eq:prob}
 P_N(e_{\rm IS},T) =  \exp\, N\,\left[s_c(e_{\rm IS}) 
                                   -\beta f_b(T,e_{\rm IS})
                               \right] / Z_N(T)
\end{equation}
therefore, in the thermodynamic limit, it will populates mainly
$e_{\rm IS}$ with energy $e(T)$ fixed by 
the condition
\begin{equation}
\label{eq:eisT}
  s_c(e) -\beta f_b(T,e) = \mbox{maximum}.
\end{equation}
and the free energy (density) of the system becomes
\begin{equation}
\label{eq:freen}
f(T) = f_b(T,e(T)) - T\, s_c(e(T)).
\end{equation}
The separation of the free energy into two  contributions reflects the
time-scale separation between inter and intra basin motions.
Condition (\ref{eq:eisT}) is equivalent to that of $f(T)$ being
minimal, i.e., 
\begin{equation}
\label{eq:minf}
\frac{\partial f}{\partial e} = \frac{\partial f_b(T,e)}{\partial e} 
                               -T\, \frac{\partial s_c(e)}{\partial e} 
                              = 0.
\end{equation}
Note that the minimum condition follows from the balance between
the contribution from the change with the energy of the shape of the basins 
($\partial f_b(T,e) / \partial e$) and 
its corresponding number ($\partial s_c(e) / \partial e$).
 
Often it is useful to write the basin free energy as
$f_b(T,e_{\rm IS}) = e_{\rm IS} + f_v(T,e_{\rm IS})$
to evidenciate the contribution from the motion inside the 
$e_{\rm IS}$ basins. Indeed from (\ref{eq:freen}) 
the average internal energy density reads 
$u(T) = \partial (\beta f_b) / \partial \beta =$
$e(T) + \partial (\beta f_v) / \partial \beta$
where 
the first term is the (average) energy of $e_{\rm IS}$ relevant 
at temperature $T$, while  the second term is the contribution 
from fluctuations inside the $e_{\rm IS}$ basin.
The contribution $f_v$ is called ``vibrational''.

\begin{figure}[hbt]
\epsfbox{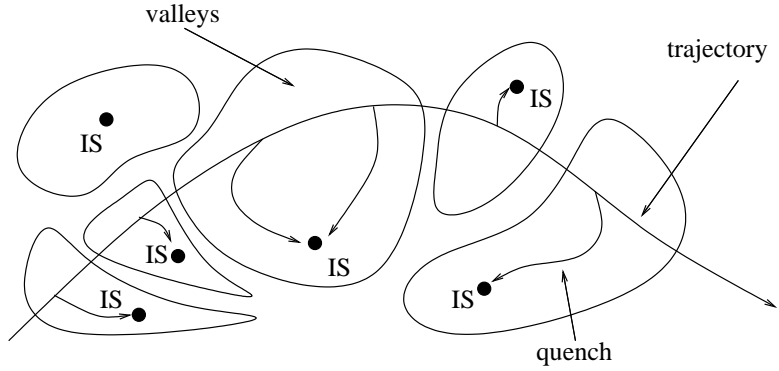}
\caption{Stillinger and Weber decomposition
}
\label{fig:sw}
\end{figure}

The main advantage of the SW decomposition is that it can be easily
transformed into a numerical algorithm, and the 
increase of the computational power has greatly pushed the
IS analysis of the energy surface.
The scheme, summarized in Figure \ref{fig:sw}, follows directly from 
the definition.

First we equilibrate the system at a given temperature $T$, 
then starting from an equilibrium configuration the system is
instantaneously quenched down to $T=0$ by decreasing the
energy along the steepest descent path. The procedure is repeated 
several times starting from uncorrelated equilibrium configurations.
In this way the IS are identified and quantities such 
as the $e_{\rm IS}$ probability distribution or $e(T)$ computed. 

\begin{figure}[hbt]
\epsfbox{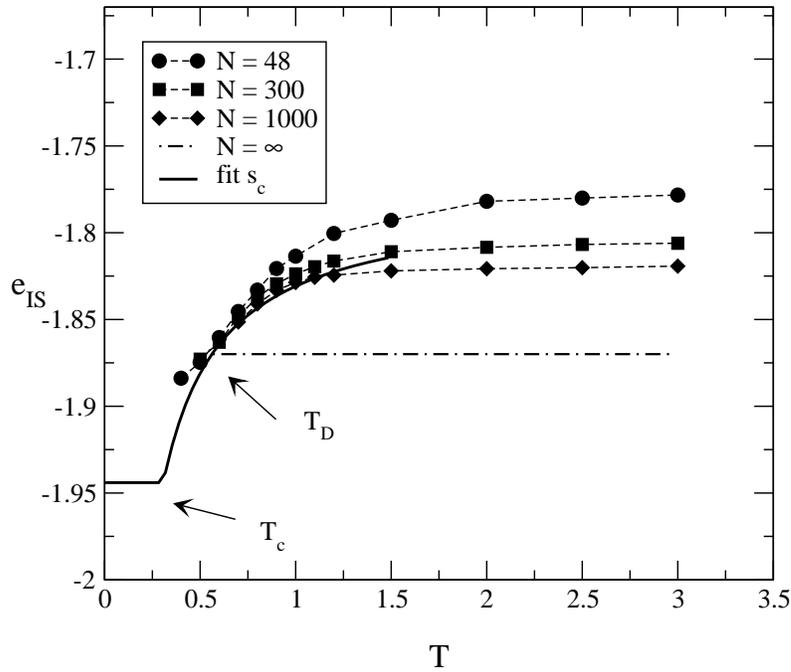}
\caption{Temperature dependence of $e(T)$ for the Random Orthogonal
         Model with 
         $N=48$ (square), $N=300$ (circle) and $N=1000$ (diamond). 
	 The horizontal line is the $N\to\infty$ limit.
         The arrows indicate the critical dynamic temperatures $T_D$
         (dynamic or Mode Coupling)   
         and $T_c$ (static or Kauzman). 
         The line is the curve obtained from the
         configurational entropy for large $N$.
	 [see also Ref. \cite{CR00a}].
}
\label{fig:eis-rom}
\end{figure}

\begin{figure}[hbt]
\epsfxsize=10cm\epsfysize=7cm
\epsfbox{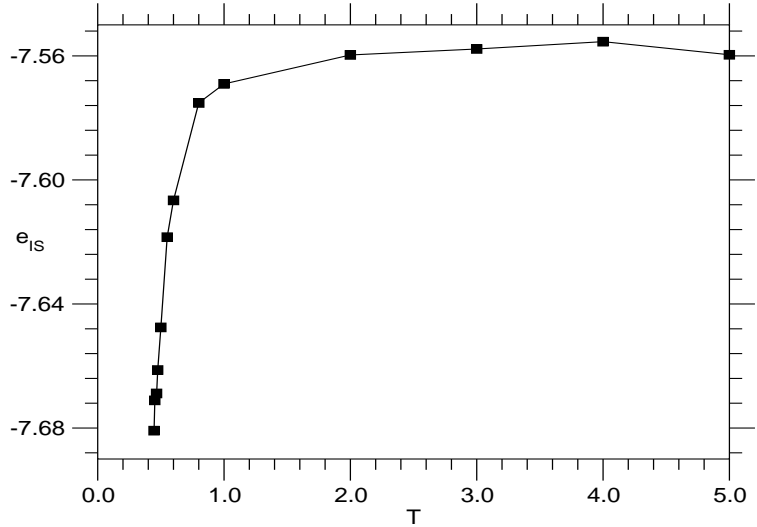}
\caption{Temperature dependence of $e(T)$ for the 
         a binary (80:20) mixture of Lennard-Jones particles. 
         Simulations were done for $1000$ particles at a fixed density 
         of $1.2$ 
         [Data courtesy of W. Kob, F. Sciortino and P. Tartaglia;
          see also Ref. \cite{KST99}].
}
\label{fig:eis-bmlj}
\end{figure}

In Figure \ref{fig:eis-rom} we report $e(T)$ as a function 
of temperature $T$ for the Random Orthogonal Model (ROM) 
of different sizes, while in Figure \ref{fig:eis-bmlj} the
same quantity is shown for a binary mixture of Lennard-Jones (BMLJ)
particles.

 From the figures we see a sharp drop 
in the $e_{\rm IS}$ energy as temperature is lowered.
For all systems, both with discrete or continuous variables, displaying a
fragile glass transition studied so far 
the drop turns out to be strongly correlated with 
the onset of slow dynamics \cite{Frag}. Indeed the decreasing of
the $e_{\rm IS}$ value is 
a clear  indication that the system explores deeper and deeper basins.

\begin{figure}[hbt]
\epsfxsize=10cm\epsfysize=8cm
\epsfbox{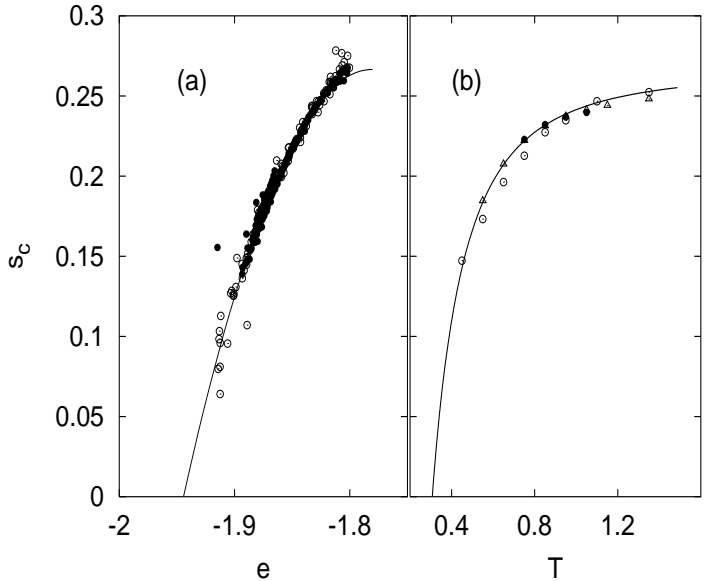}
\caption{(a) Configurational entropy as a function of energy for the ROM.
         The data are from system sizes $N=48$ (empty circle) and $N=300$
         (filled circle), and temperatures $T=0.4$, $0.5$, $0.6$, $0.7$, 
         $0.8$, $0.9$ and $1.0$. For each curve the unknown constant has
         been fixed to maximize the overlap between the data and
         the theoretical result \protect\cite{ROM}. The line is the
	 quadratic best-fit.
         (b) Configurational entropy density as a function of temperature.
         The line is the result from the best-fit of $s_c(e)$ 
         while the symbols are the results from the temperature integration
         of eq. (\protect\ref{eq:entro}) for $N=48$ (empty circle), 
         $N=300$ (empty triangle) and $N=1000$ (filled circle). 
	 [see also Ref. \cite{CR00a}].
}
\label{fig:compl}
\end{figure}

 From the knowledge of $e_{\rm IS}$ distribution we can reconstruct the
SW complexity $s_c(e)$ simply inverting eq. (\ref{eq:prob}):
\begin{equation}
\label{eq:compl}
  s_c(e) = \ln P_N(e,T) + \beta e + \beta f_v(T,e) +
               \ln Z_N(T)
\end{equation}
If energy dependence of $f_v(T,e)$ can be neglected, then it is possible 
to superimpose the curves for different temperatures. 
The resulting curve is, except for
an unknown constant, the SW complexity $s_c(e)$. 
This is shown for the ROM in Figure \ref{fig:compl} panel (a). 
The data collapse is rather good for $e < -1.8$,
indicating that above the energy dependence of $f_v(T,e)$ cannot be neglected.
The line is the quadratic best 
fit which gives the value $e_c\simeq -1.944$
for the critical energy 
where $s_c(e)$ vanishes  in good agreement with the theoretical 
result $e_{1rsb}=-1.936$ \cite{ROM}.

Direct consequence of $f_v(T,e) \simeq f_v(T)$ for $e<-1.8$ is that
it drops out from eq. (\ref{eq:minf}) so that the minimum condition simply
reads:
\begin{equation}
\label{eq:entro}
 \frac{d\, s_c(e)}{d\, e} = \frac{1}{T}.
\end{equation}
 From this relation, by integrating the $T$ dependence of $d\, e/T$,
we can compute $s_c$ as a function of $T$. The result obtained 
using the data of Figure \ref{fig:eis-rom} is shown in 
Figure \ref{fig:compl} panel (b). The line is the result valid for large 
$N$ obtained using the quadratic best fit of panel (a).

As discussed above the vibrational contribution $f_v$
follows from the motion inside the $e_{\rm IS}$ basins. Its independence
from $e_{\rm IS}$ means that all basins are equivalent, i.e., have
the same shape. In general this is not the case and its contribution
must be included. For systems with continuous variables, as for example 
BMLJ particles, at low $T$ 
$f_v$ can be calculated in the harmonic approximation by expanding the
energy about the IS configuration:
\begin{equation}
f_v(T, e_{IS}) = k_B T \sum_{i=1}^{3N-3} 
\ln\,\left[\hbar \omega_i(e_{\rm IS})/k_B T \right]
\label{eq:hfvib}
\end{equation}
where $\omega_i(e_{\rm IS})$ is the frequency of the $i$-th normal mode
in the $e_{\rm IS}$ basin which in general depends on the specific 
IS configuration, i.e., different IS with the same $e_{\rm IS}$ may 
have different normal modes.
If all basins had the same curvature, then $f_v$ would be only
function of $T$ and we are back to the previous case.
In the BMLJ system, basins with different $e_{\rm IS}$ have
different curvatures \cite{ST01} and hence, at difference with the ROM, 
 $f_v$ is a function of both $T$ and $e_{\rm IS}$.

\subsection{Violation of FDT and Effective Temperature}

More informations on the structure of the energy surface 
can be obtained from non-equilibrium relaxation processes.  
We shall consider here the non-equilibrium behaviour of the
system following an instantaneous quench from an equilibrium state 
a temperature $T_i$ above the glass transition $T_g$ to a temperature
$T_f$ below it.

For temperature close to the mode coupling critical temperature, where
the intra-basin and inter-basin time scales become well separated, the
vibrational intra-basin dynamics quickly thermalize to the thermal 
bath temperature $T_f$. The thermalization of the entire
system is instead rather slow, being dominated by the inter-valley
processes. Therefore the fast equilibration of the
intra-basin degrees of freedom is followed by a much slower
process during which the system populates deeper and deeper $e_{IS}$ levels.

\begin{figure}[hbt]
\epsfxsize=10cm\epsfysize=8cm
\epsfbox{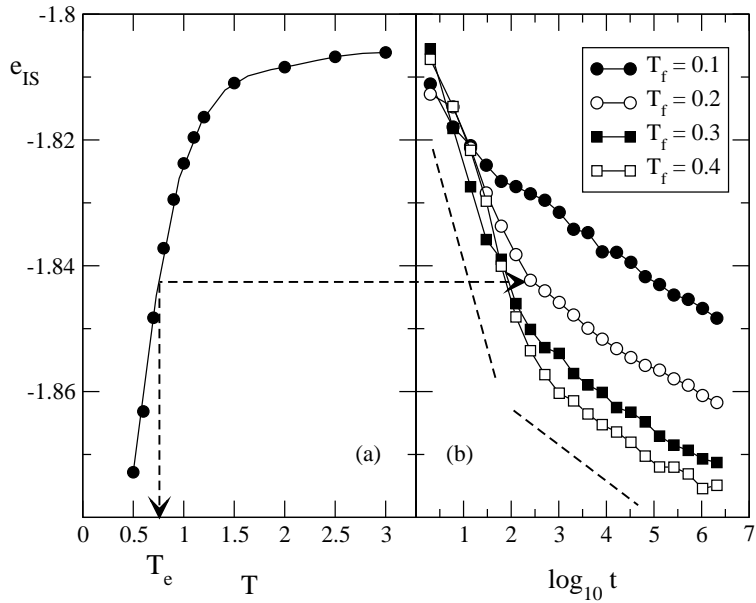}
\caption{Panel (b): 
         Average inherent structure energy for the ROM as function of time for
         initial equilibrium temperatures $T_i= 3.0$ and final
         temperatures $T_f= 0.1$, $0.2$, $0.3$ and $0.4$. The average is over
         $300$ initial configurations. The system size is $N=300$. 
         The lines denotes the two regimes.
         Panel (a): Equilibrium average $e_{\rm IS}$ a function of 
         temperature. The arrows indicate the construction of the
         effective temperature $T_e(e_{\rm IS})$.
	 [see also Ref. \cite{CR00b}].
}
\label{fig:e-agi-rom}
\end{figure}

In the right panel of Figure \ref{fig:e-agi-rom} we show the 
average $e_{\rm IS}$ energy 
as a function of time after the quench for the ROM. 
The figure reveals that the relaxation process can be 
divided into two different regimes. A first regime
with a power law 
independent of $T_f$, and a second regime 
with a power law independent of 
both $T_i$ and $T_f$. The final temperature $T_f$ controls the cross-over 
between the two regimes. A similar behavior has been observed in 
molecular dynamics simulations of super-cooled liquids \cite{KST99}. 

The two regimes are associated with different relaxation processes.
In the first part the system has enough energy and
relaxation is mainly due to {\it path search} out of basins through
saddles of energy lower than $k_{B}T_f$.
This part depends only on the initial 
equilibrium temperature $T_i$ and should slow down as 
$T_i$ decreases since, as reasonable, lower states are surrounded by higher 
barriers.
This process stops when all barrier heights
become of $O(k_{B}T_f)$ and relaxation slows down since it must
proceed only via activated inter-valley processes.

\begin{figure}[hbt]
\epsfxsize=10cm\epsfysize=8cm
\epsfbox{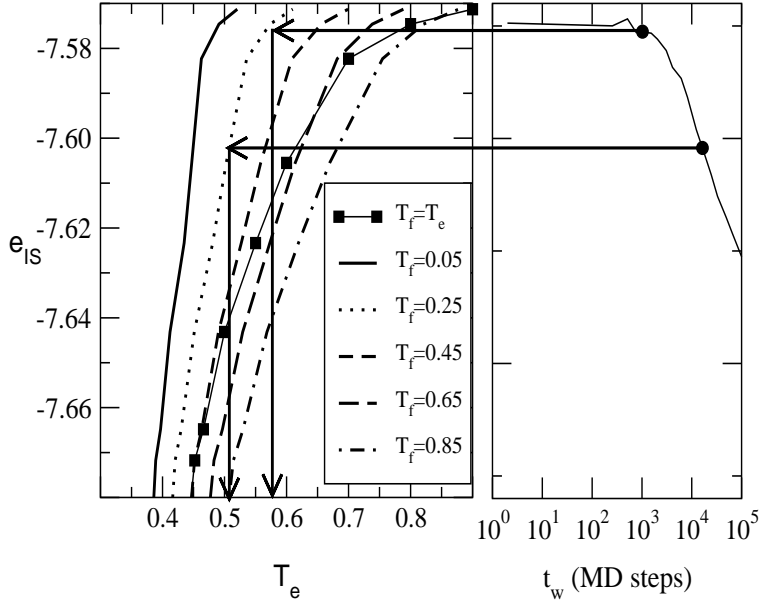}
\caption{Left: Solutions of  Eq.~(\ref{eq:teff1}) for several  values of $T_f$ 
         for the BMLJ system. 
         Right: $e_{\rm IS}$ as a function of time, following the 
         temperature quench.
         The arrows show graphically the procedure which connects 
         the $e_{\rm IS}(t)$ value to $T_e$ value, once $T_f$ is known. 
         [Data courtesy of F. Sciortino and P. Tartaglia,
          see also Ref. \cite{ST01}].
}
\label{fig:teff-bmlj}
\end{figure}

Under the assumption of a fast equilibration of the intra-basin motion,
we can define an {\it effective temperature} $T_e$ as the temperature
the system would have when populating the basins of depth $e_{\rm IS}$.
The temperature $T_e$ can be obtained from (\ref{eq:minf}) with 
the vibrational contribution evaluated
at the bath temperature $T_f$, since we assume local equilibrium of
the intra-basin motion, and resolving for $T$ \cite{ST01}:
\begin{equation}
\label{eq:teff1}
T_e(e_{\rm IS},T_f) = \frac
                        {1+ (\partial /  
                             \partial e_{\rm IS})\, f_v (T_f,e_{\rm IS})
                             }
                             { (\partial /
                                \partial e_{\rm IS})\,   s_c(e_{\rm IS}) 
                             }
\end{equation}
The effective temperature $T_e$ for the BMLJ is shown in
in Figure \ref{fig:teff-bmlj}.
If $f_v$ is not a function of $e_{\rm IS}$, as 
in the case of ROM, the above equation reduces to \cite{CR00a,CR00b,CR00}
\begin{equation}
\label{eq:teff2}
T_e(e_{\rm IS}) ^ {-1} =  \frac{\partial s_c(e_{\rm IS})}{\partial e_{\rm IS}}
\end{equation}
and curves with different $T_f$ coincide, see Figure \ref{fig:e-agi-rom}.
In the non-equilibrium relaxation process the value of $e_{\rm IS}$ will
vary with time making $T_e$ a function of time, since the r.h.s.
of eqs. (\ref{eq:teff1}) and (\ref{eq:teff2}) must be evaluated at 
$e_{\rm IS}(t)$.  For each time $t$ the
value of $T_e(t)$ can then be obtained graphically as shown in
Figures \ref{fig:e-agi-rom} and \ref{fig:teff-bmlj}.

The predictions of the quasi-equilibrium assumption can be tested by 
studying the response of the system after a quench to $T_f$ 
at time $t=0$ 
to a perturbation switched on at some later time $t_w$.
In the linear response regime the average value of any observable $A$
at time $t$ after a perturbation field $h_A$ conjugated to it 
is applied to the system at $t_w<t$ is 
$\langle A(t) \rangle = \chi_{\rm ZFC}(t,t_w)\, h_A$, where the
Zero Field Cooled susceptibility $\chi_{\rm ZFC}(t,t_w)$ is
related to the $AA$ correlation function through the famous
Fluctuation-Dissipation formula \cite{Note}
\begin{equation}
\chi_{\rm ZFC}(t,t_w) =
 \left.  \frac{1}{T}\, 
   \Bigl[ \langle A(t)\,A(t)\rangle - \langle A(t)\,A(t_w)]\rangle \Bigr]
 \right|_{h_A=0}.
\end{equation}
which predicts that the response be proportional to $T^{-1}$.
Under the assumption that the intra-valley motion quickly 
thermalizes to the thermal bath temperature $T_f$, while the 
thermalization of the entire system is dominated by the 
slow inter-valley processes, it follows that
for short times $T=T_f$ since the intra-basin motion is probed, while
for long times $T=T_e$ since now is the inter-basin to be probed.
The first regime corresponds to the region where the correlation function 
assumes values between the equal time
and the plateau values, while the second where it assumes values below the 
plateau value.

\begin{figure}[hbt]
\epsfxsize=10cm\epsfysize=8cm
\epsfbox{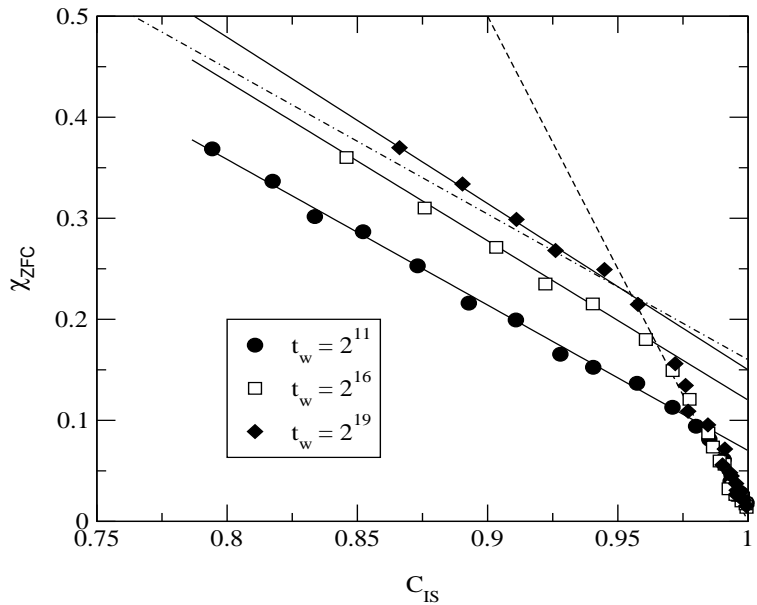}
\caption{Integrated response function as a function of IS correlation
         function, i.e, the correlation between different IS configurations,
         for the ROM.  The dash line has slope $T_f^{-1} = 5.0$, 
         while the full lines are the prediction
         (\protect\ref{eq:teff2}):
         $T_e(2^{11})\simeq 0.694$, 
         $T_e(2^{16})\simeq 0.634$ and
         $T_e(2^{19})\simeq 0.608$.
         The dot-dashed line is $T_e$ for
         $t_{\rm w}=2^{11}$ drawn for comparison.
	 [see also Ref. \cite{CR00b}].
}
\label{fig:fdt-rom}
\end{figure}

\begin{figure}[hbt]
\epsfxsize=10cm\epsfysize=8cm
\epsfbox{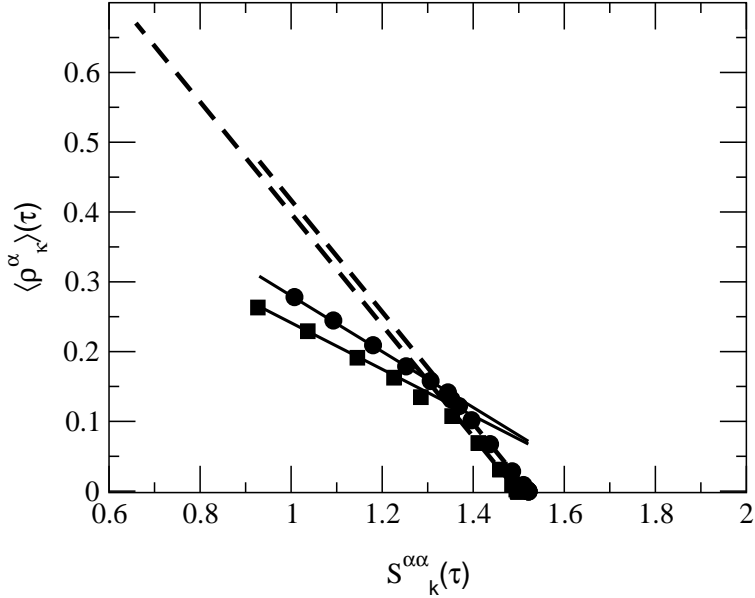}
\caption{Response $\langle  \rho^{\alpha}_{\bf k}(\tau)\rangle$ 
         versus the dynamical structure factor 
         $S_{\bf k}^{\alpha\alpha}(t) \equiv \langle \rho^{\alpha}_{\bf k}(t)
         \rho^{\alpha*}_{\bf k}(0) \rangle$, where 
         $\rho^{\alpha}_{\bf k}$ is the Fourier transform component of 
         the density of $\alpha=A,B$ particles at wave-vector ${\bf k}$,
         for the Binary Mixture Lennard-Jones 
         particles system for $T_i = 0.8$, $T_f=0.25$ and two 
         waiting times $t_w=1024$ (square) and $t_w=16384$ (circle).
         Dashed lines have slope $T_f^{-1}$ while thick lines have slope
         $T_e^{-1}$. 
         [Data courtesy of F. Sciortino and P. Tartaglia, 
          see also Ref. \cite{ST01}].
}
\label{fig:fdt-bmlj}
\end{figure}

In Figure \ref{fig:fdt-rom} we show the response versus correlation
plot for the ROM. The analogous plot for the BMLJ is in Figure
\ref{fig:fdt-bmlj}. As follows from both figures for short times
the plot is linear with the
expected $T_f^{-1}$ slope, properly describing the equilibrium
condition of the intra-basin dynamics with the external bath.
At larger times, the inter-basin motion sets-in and the slope
becomes $T_e^{-1}$,   in very good agreement with the value
predicted by eq.~(\ref{eq:teff1}).

\section{Beyond the Stillinger and Weber projection: the free 
         energy landscape}

In the previous sections the main effort has been to characterize glassy
dynamics by studying the structure of minima of the potential energy
landscape. 
Nevertheless, as has been already said valleys
are not only characterized by the energy at their bottom but also by the
size of the basins of attraction.
The decomposition proposed by
Stillinger and Weber is meaningful if the typical time to explore a
given IS or valley only depends on the energy of that valley (defined,
for instance by the energy $e_{\rm IS}$ of its associated minimum). 
But one can
imagine a situation where IS with the same energy have completely
different basins of attraction. In that case the probability to explore a
given IS not only depends on its energy but also on the size of the basin of
attraction or its associated entropy. The most natural approach in which these
considerations are properly taken into account is to assume that every IS
is characterized by its free energy $F_{\rm IS}(T)$ defined as:
\begin{equation}
\exp(-\beta F_{\rm IS}(T))=\sum_{{\cal C}\in IS}\exp(-\beta E({\cal C})) 
\label{free1}
\end{equation} which corresponds to the free energy of 
a portion of the whole phase
space  containing all configurations belonging to the specific IS. We
can now extend \cite{Rocco} 
the ideas of Stillinger and Weber to include the free
energy in the formulation by using the following equiprobability
assumption: when the system is in equilibrium at a given temperature,
valleys with the same free energy have the same probability to be
explored. Both the number
of configurations contained in each valley and the number
$g(F,T)$ of valleys with a given free energy 
grow exponentially with the size of the system
leading to a proper thermodynamic formulation in the large volume
limit. The equilibrium partition function can be written in terms
of the IS free energies (\ref{free1}) as:

\begin{eqnarray}
{\cal Z} &=& \sum_{\cal C}\exp(-\beta E({\cal C}))
          =\sum_{\rm IS} \sum_{{\cal C}\in {\rm IS}}\exp(-\beta E({\cal C}))
\nonumber\\
          &=&\sum_{\rm IS}\exp(-\beta F_{\rm IS}(T))
\label{free2}
\end{eqnarray}
At a given
temperature, the average free energy among all valleys is determined
by a balance between the probability to explore
valleys with free $F$ (proportional to the Boltzmann factor $\exp(-\beta F)$) 
and the number $g(F,T)$ of valleys with
that free energy. Hence, the equilibrium free energy is given
by 

\begin{eqnarray} 
   \exp(-\beta F_{\rm eq}) &=& \sum_{\rm IS}\exp(-\beta F_{\rm IS}(T))
           =\sum_Fg(F,T)\exp(-\beta F)
\nonumber\\
           &=&\sum_F \exp(S_c(F,T)-\beta F)=\sum_F \exp(-\beta\Phi(F,T))
\label{free3}
\end{eqnarray}
where $S_c(F,T)=\log(g(F,T))$ defines the configurational entropy 
while the
function $\Phi(F,T)=F-TS_c(F,T)$ is the thermodynamic potential
associated to it. Because $\Phi$ is an
extensive quantity, in the large-volume limit, the leading contribution
to (\ref{free3}) is determined by the minimum of $\Phi(F,T)$ as function
of $F$,

\begin{eqnarray}
F_{\rm eq}(T)=\Phi(F^*,T)=F^*-TS_c(F^*,T)\label{free4}\\
\frac{1}{T}=\left.\frac{\partial S_c(F,T)}{\partial F}\right|_{F=F^*}~~~.
\label{free5}
\end{eqnarray}
Note that the average free energy $F^*$ does not coincide
with the equilibrium free energy $F_{\rm eq}(T)$ but it is always
higher, $F^*=F_{\rm eq}(T)+TS_c(F^*,T)$, the difference being the
configurational entropy evaluated at $F^*$.
 
It is interesting to note the parallelism between this coarse-grained
description and the standard equilibrium theory. In that case, the
relevant entities are the configurations and the equilibrium energy is
given by a balance between the Boltzmann factor $\exp(-\beta E)$ and
the degeneration $g(E)=\exp(S(E))$ where $S(E)$ is the entropy. The
relation $1/T=\frac{\partial S(E)}{\partial E}$ yields the
thermodynamics.  Nevertheless, an important difference between the
standard equilibrium theory and the present free energy valley
decomposition must be stressed. Both the equilibrium entropy $S(E)$ and
the Stillinger and Weber configurational entropy $S_c(E)$ are only
functions on the energy while the configurational entropy $S_c(F,T)$ in
this new formalism depends on both the free energy and the temperature. 
Therefore the configurational entropy turns out to be a more complicated
object when expressed in terms of the free energy than in terms of the
energy of the valleys.

A natural question arises now: what are the assumptions behind the
validity of this free energy decomposition?  The main assumption in
Boltzmann theory is the equiprobability assumption, i.e., all
configurations with identical energy have in equilibrium 
the same probability.
This introduces the equilibrium measure used in ensemble
theory which is at the roots of Statistical Mechanics. For glassy
systems a similar idea is behind the physical meaning of the present
free energy decomposition. One assumes \cite{Rocco} the validity of an
equiprobability hypothesis stating that valleys with identical free
energy have the same probability to be explored. This gives a flat
measure in equilibrium which, if extended to the non-equilibrium case,
is the analogous to the Edwards measure proposed in the context
granular media \cite{EDME,MOPO,Selli}.  The main difference between the
Edwards measure and this new free energy measure is that the former
occurs in a non stationary sheared situation \cite{KURCHAN} or even in
an stationary one (for instance, under tapping \cite{DEAN}) at zero
temperature while the latter occurs in a non-stationary relaxational
regime at finite temperature.

In a recent work \cite{Rocco} the validity of this free energy measure
 has been explicitly tested by studying the configurational entropy of
 simple models. This has been done introducing a probabilistic argument
 to compute the free energy of the IS. In equilibrium, the probability
 to explore a given IS, is given by $p_{\rm IS}(T)=\exp(-\beta(F_{\rm
 IS}(T)-F_{\rm eq}(T)))$. Then one can run a simulation and, after
 equilibrating, collect the number of times $N_{\rm IS}$ that a
particular IS is
 found among a total number of quenches $N_{\rm run}$. This yields
 $p_{\rm IS}(T)=N_{\rm IS}/N_{\rm run}$ from which we have an estimate of the
 IS free energy: 
$F_{\rm IS}(T)=-T\log(p_{\rm IS}(T))+F_{\rm eq}(T)$. 
By using this method the configurational entropy $S_c(F,T)$ has 
been computed in Ref. \cite{Rocco} for the ROM and the Sherrington 
and Kirkpatrick model, a model 
with a completely different energy surface topology \cite{Usua}.
The average free energy of the IS, $F^*$ can be obtained 
from the minimum of the potential $\Phi(F,T)$. 
The difference between the minimum
$F^*(T)$ and the equilibrium free energy $F_{\rm eq}(T)$ yields the
 configurational entropy at the given temperature $S_c(F^*,T)$. 
Moreover from the shape of $\Phi(F,T)$ it is possible to infer
both the type of transition of the model, one-step or infinite-replica
symmetry breaking, as well as the critical o Kauzman temperature.
We note that as
it happens for the Stillinger and Weber results discussed in
 the previous section, the method works only for finite-sized
 systems where the number of different IS is not too large.

The generalization of this free-energy scenario to the dynamics has been
proposed in \cite{HETERO} where it has been proposed that, the effective
temperature in structural glasses is related to its fragility.
Moreover, the effective temperature is also given by the slope of the
configurational entropy evaluated at the threshold free energy. This
off-equilibrium scenario complements the new measure discussed
before and offers a scenario for the glass transition driven by entropic
barriers. 

\section{Models with kinetical constraints.}

Another class of interesting models are kinetically constrained
models. These models are characterized by a extremely simple
thermodynamic behavior without any type of phase transition but with a
complicated slow dynamics due the existence of kinetical constraints.
The constraints are such that detailed balance and ergodicity are
preserved despite the infinite energy barriers they introduce.
In
some sense they are similar to hard spheres models where some
configurations are excluded form the configurational space although
with much simpler static properties. The simplest example of these
family of models is the kinetically constrained Ising paramagnet where
there is no interaction between the spins (hence there is no critical
point even at $T=0$) but some transitions are excluded in the
dynamics.  The description of these models within an IS formalism has
been studied in detail in \cite{ISno}. Here we only want to make some
general considerations about the validity of the IS description of the
dynamics, for a more detailed discussion we refer to Ref. \cite{ISno}.

In general, in most of the dynamically constrained models the energy
function can take only a discrete set of values, therefore there is
accidental degeneracy in the density of states. One of the main
problems related to this fact and discussed in Ref. \cite{ISno} is
that it is difficult to properly define the IS decomposition
itself. The steepest descent procedure to map a configuration onto a
valley is not well defined because it is not univocally defined. From
a physical point of view this does not seem to be a serious
problem. Suppose we add to the original Hamiltonian a random
perturbation term $P$ which lifts the degeneration. If the system is
stochastically stable (the dynamical behavior in the limit $P\to 0$
coincides with the dynamical behavior of the unperturbed system) then
one can work out with the perturbed system and make the perturbation
vanish at a later stage. Because the dynamics of the vast majority of
glassy systems is probably stochastically stable we believe that this is not
serious problem of the approach.

But there is another problem which seriously compromises the validity
of the IS approach. One can show that the IS decomposition is
completely identical for some models with completely different
dynamical constraints, so that the SW configurational entropy is
exactly the same.  Because the dynamics of these models is known to be
extremely different it is clear that such configurational entropy
cannot describe their relaxational dynamics.  Obviously, to cope with
this problem one can go further and describe the dynamical behavior in
terms of the configurational entropy but now defined in terms of the
free energy. This route could eventually solve the problem because
now, although the energy of the different IS is identical for all
models, the entropic contribution (as explained in the preceding
section) can be completely different.  Nevertheless, we believe that
going beyond the standard SW description will not really solve another
more essential problem present in this kind of systems. One of the
main features of these models is that relaxation occurs by the
coarsening of a typical length scale. This length scale reflects the
typical size of spatial regions which are ordered into the ground
state structure. As relaxation proceeds the configuration of the
system approaches that of the ground state. Now, the ground state in
this class of systems corresponds to the crystal structure in
structural glasses. So the relevant question is: How important is the
existence of a crystalline configuration in structural glasses as far
as the glass transition phenomenology is concerned? Light scattering
measurements point into the direction that relaxation in the
under-cooled region does not proceed by crystallization of larger and
larger regions and that the structure of the glass is always that of a
liquid. This result is in agreement with the very well known fact that
some spin glasses with disorder (where there does not exist a
crystalline structure) display a behavior similar to structural
glasses. Hence we must conclude that the slow dynamics in glasses is
not necessarily related to a coarsening process into a single and
unique structure. The physical mechanism behind relaxation then must
be some kind of entropic search thermally activated but without any
kind of growing order in the system. On the other hand the existence of a
some kind of coarsening process in the relaxation 
will preclude the validity of the equiprobability hypothesis, and
hence the existence of a flat measure for the free energy.
Therefore kinetically constrained models, despite their striking
non-equilibrium behavior very similar to that of glasses, cannot be
described in a framework as that proposed in the previous
section.  Nevertheless, a careful study of the constrained Ising chain
in terms of the free energy could be certainly interesting to better
elucidate this question.

\section{Conclusions}

There is still more work to be done. Most of the lines of research here presented 
will clearly see a fast development in the next future when
our understanding of the validity of general scenarios to describe
glassy systems will improve. The most important results we have tried
to propose in this paper can be summarized in the form of answers 
to the following set of selected questions,

\begin{itemize}

\item{\em When the SW decomposition is expected to work fairly well?}
A statistical description of valleys in terms of inherent structures
seems to be useful for systems where valleys are very narrow
(i.e. containing a exponentially large number of configurations
scaling like $\exp(\alpha N)$ but with $\alpha\ll 1$ to assure that
the entropic contribution to the free energy valley is small) or
valleys very large but with very similar shape. The later assumption
corresponds to the claim that the distribution of instantaneous
frequency modes computed in the harmonic approximation for those
valleys are not too broad functions.  This is indeed the case for,
e.g., BMLJ \cite{Sciop}. Obviously, glassy systems where there is a
highly heterogeneous distribution of basins of attraction for the
inherent structures cannot be described within the usual SW approach.

\item{\em How one can improve the standard SW approach?} The idea to
include the basins of attraction in the dynamical formulation is by
supposing that basins are visited according to their free
energy. Therefore, in equilibrium at temperature $T$, the probability
to visit a valley with free energy $F$ is proportional to the Boltzmann
factor $\exp(-\beta F)$ and to their number 
$g(F,T)=\exp(S_c(F,t))$ where $S_c(F,T)$ defines the configurational
entropy. This free-energy measure can be further extended to deal with
non-equilibrium processes where the probability to jump among valleys
is simply given by the entropic term $g(F^*,T)$ evaluated at the
time-dependent threshold free energy $F^*=F(t)$.  This description
offers an interpretation of the violation of $FDT$ in terms of a
single timescale given by the effective temperature evaluated at the
threshold $1/T_{\rm e}(t)=(\frac{\partial S_c(F,T)}{\partial
F})_{F=F^*}$. The validity of this flat measure in terms of the free
energy in the off-equilibrium regime remains one of the most
fascinating problems when trying to have a general theory for slowly relaxing 
non-equilibrium processes in complex free energy landscapes.

\item{\em What is the utility of investigating relatively small
systems?} As we have stressed in the previous item a description of
basins in glassy systems must be done in terms of the free-energy
landscape.  The appropriate configurational entropy is then a function
of both the free energy and temperature. Within the IS formalism an
estimate of the free energy of each IS can be done by sampling the
IS-space. A good sampling requires that each IS is visited with a
finite frequency. If the number of IS is too large this is not
possible. Considering that the total number of IS is exponentially
large with the volume of the system we conclude that sizes must be
modest for the procedure to be implemented. Moreover, in the case of
mean-field models, one can do careful checks of the main theoretical
assumptions by comparing numerical results with analytical
ones. Furthermore, a theoretical analysis of finite $N$ corrections in
mean-field systems could give a theoretical framework for activated
processes in glassy systems.

\item{\em Is the IS formalism relevant for coarsening models?}  In
principle, for systems where a given pattern grows with time the
dynamics cannot be expressed in terms of jumps among uncorrelated
structures. Therefore, the entropic assumption is not justified and a
dynamical measure in terms of the free energy landscape does not hold
anymore. In this respect it would be extremely interesting to find a
coarsening model where the IS formalism in terms of the free energy
works. To our knowledge, such an example has not been yet provided.

\end{itemize}

In summary, the IS description of dynamics in terms of the energy and,
more generally, in terms of the free energy of basins provides the
first approximate scheme to deal with the dynamics of complex
systems. There are still obscure and not well understood points which
hopefully will be progressively clarified in the near future. This
will provide a more complete comprehension of the main physical
mechanisms behind the elusive glass transition problem.

\ack We thank for useful discussions B. Coluzzi, C. Donati, U. Marini
Bettolo, E. Marinari, A. Rocco F. Sciortino, M. Sellitto and
P. Tartaglia.  A special thank goes to F. Sciortino for having given
us the data for BMLJ systems.  F.R. acknowledges funding from the
project PB97-0971 and Acciones Integradas collaboration HI2000-0087.


\end{document}